\input{aipcheck}

\documentclass[
    ,final            
  ]
  {aipproc}

\layoutstyle{6x9}
\usepackage[rightcaption]{sidecap}

\newcommand{\ga}{\alpha}
\newcommand{\gb}{\beta}

\begin{document}

\title{Dynamics of few-body states in a medium}

\author{M.~Beyer}{
  address={Fachbereich Physik, Universit\"at Rostock, D-18051
  Rostock, Germany} 
}

\author{S.~Mattiello}{
  address={Fachbereich Physik, Universit\"at Rostock, D-18051
  Rostock, Germany} 
}

\author{S.~Strau\ss}{
  address={Fachbereich Physik, Universit\"at Rostock, D-18051
  Rostock, Germany}
}

\author{T.~Frederico}{
  address={Dep. de Fisica, ITA, CTA, 12.228-900 S\~ao Jos\'e dos Campos, S\~ao Paulo, Brazil}
}
\author{H.~J.~Weber}{
  address={Dept. of Physics, University of Virginia, Charlottesville VA, U.S.A.}
}

\author{P.~Schuck}{
  address={Institut de Physique Nucl\'eaire, F-91406, Orsay Cedex, France}
}

\author{S.A.~Sofianos} {
  address={Physics Department, University of South Africa, Pretoria 0003,
  South Africa}
}

\begin{abstract}
  Strongly interacting matter such as nuclear or quark matter leads to
  few-body bound states and correlations of the constituents. As a consequence
  quantum chromodynamics has a rich phase structure with spontaneous symmetry
  breaking, superconductivity, condensates of different kinds. All this
  appears in many astrophysical scenarios. Among them is the formation of
  hadrons during the early stage of the Universe, the structure of a neutron
  star, the formation of nuclei during a supernova explosion.  Some of these
  extreme conditions can be simulated in heavy ion colliders.  To treat such a
  hot and dense system we use the Green function formalism of many-body
  theory.  It turns out that a systematic Dyson expansion of the Green
  functions leads to modified few-body equations that are capable to describe
  phase transitions, condensates, cluster formation and more. These equations
  include self energy corrections and Pauli blocking. We apply this method to
  nonrelativistic and relativistic matter. The latter one is treated on the
  light front. Because of the medium and the inevitable truncation of space,
  the few-body dynamics and states depend on the thermodynamic parameters of
  the medium.

\end{abstract}

\maketitle


\paragraph{Many-body Green functions}
To treat hot and dense quantum systems, such as nuclear or quark matter in
question here, we use the techniques of many-body Green
functions~\cite{fet71}. We organize the Green functions into a Dyson expansion
that leads to a hierarchy of linked cluster equations~\cite{duk98}.  In recent
years we have systematically applied this approach for finite temperatures up
to multi-nucleon clusters (see e.g.~\cite{Beyer:1996rx} and refs therein) and
also generalized it to light front
dynamics~\cite{Beyer:2001bc,Mattiello:2001vq}.  In fact, the light front
quantization is well suited to use the many-body Green function formalism as
in the relativistic case, as it utilizes the Fock space
expansion~\cite{Brodsky:1997de}. We define a chronological Green function
\begin{equation} 
      {\cal G}_{\alpha\beta}^{t-t'}
=-i\left(\theta(t-t')\langle A_\gb^\dagger(t')A_\ga(t)
      \rangle\mp\theta(t'-t)\langle A_\gb^\dagger(t')A_\ga(t)
       \rangle\right).\label{eqn:green}
\end{equation}
The average $\langle\cdots\rangle$ is taken over the exact ground state and
the upper (lower) sign is for fermions (bosons). The operators
$A(t)=e^{itH}Ae^{-itH}$ could be build out of any number of field operators
(fermions and/or bosons).  In finite temperature formalism the above
definition can be generalized.  For a grand canonical ensemble on the light
front, the generalized Heisenberg picture assumes the form $A(\tau)=e^{\tau
  H}Ae^{-\tau H}$, where $H=u\cdot P-\mu N$. Here $u$, $u^2=1$ is the
four-velocity of the medium. For a nonrelativistic medium at rest $u^0=1,\vec
u=0$, hence $H=P_0-\mu N$ with the Hamiltonian $P_0$, suffice. On the light
front $u^\mu=(u^+,u^-,\vec u_\perp)=(1,1,0,0)$~\cite{Beyer:2001bc}. The
ensemble average is  taken over the (equilibrium) grand canonical
statistical operator $\rho_G=Z_G/{\rm Tr} Z_G$, viz.  $\langle\cdots\rangle={\rm
  tr}\{\rho_G\dots\}$ and $$Z_G =  \exp\left\{-(u\cdot P  - \mu N)/T\right\}.$$
Dyson equations can be established for real time and imaginary time Green
functions~\cite{duk98}, also in light-front
form~\cite{Beyer:2001bc}. Neglecting memory or retardation terms the
equation for the Green function is given by
\begin{equation} 
      \left(  i\frac{\partial}{\partial{t}}\;
-  {{\cal M}^{{t}}_{0}}\right){\cal G}^{{t}-{t'}}
        =\delta({t}-{t'}) {\cal N}^{t},\label{eqn:mf}
\end{equation}
where ${\cal M}^t_{\ga\gb}=\langle
[[A_\ga,H]({t}),A_\gb^\dagger({t})]_\pm\rangle$, ${\cal
  N}^{t}_{\ga\gb}=\langle[A_\ga,A_\gb^\dagger]_\pm({t})\rangle$. Formally,
this equation holds for the nonrelativistic case and in light front form, with
$t$ replaced by the light front time
$x^+=t+z$~\cite{Beyer:2001bc,Mattiello:2001vq}. Now it is possible by
choosing the appropriate Fock operators, e.g. $A_\ga=a_1a_2a_3$ for a
three-body correlation, to derive from (\ref{eqn:mf}) cluster mean-field
equations for the few-body system embedded in a
medium~\cite{Beyer:1996rx,Beyer:2001bc,Mattiello:2001vq}.

\begin{figure}[t]
\begin{minipage}{0.5\textwidth}
\centering
\includegraphics[width=0.95\textwidth]{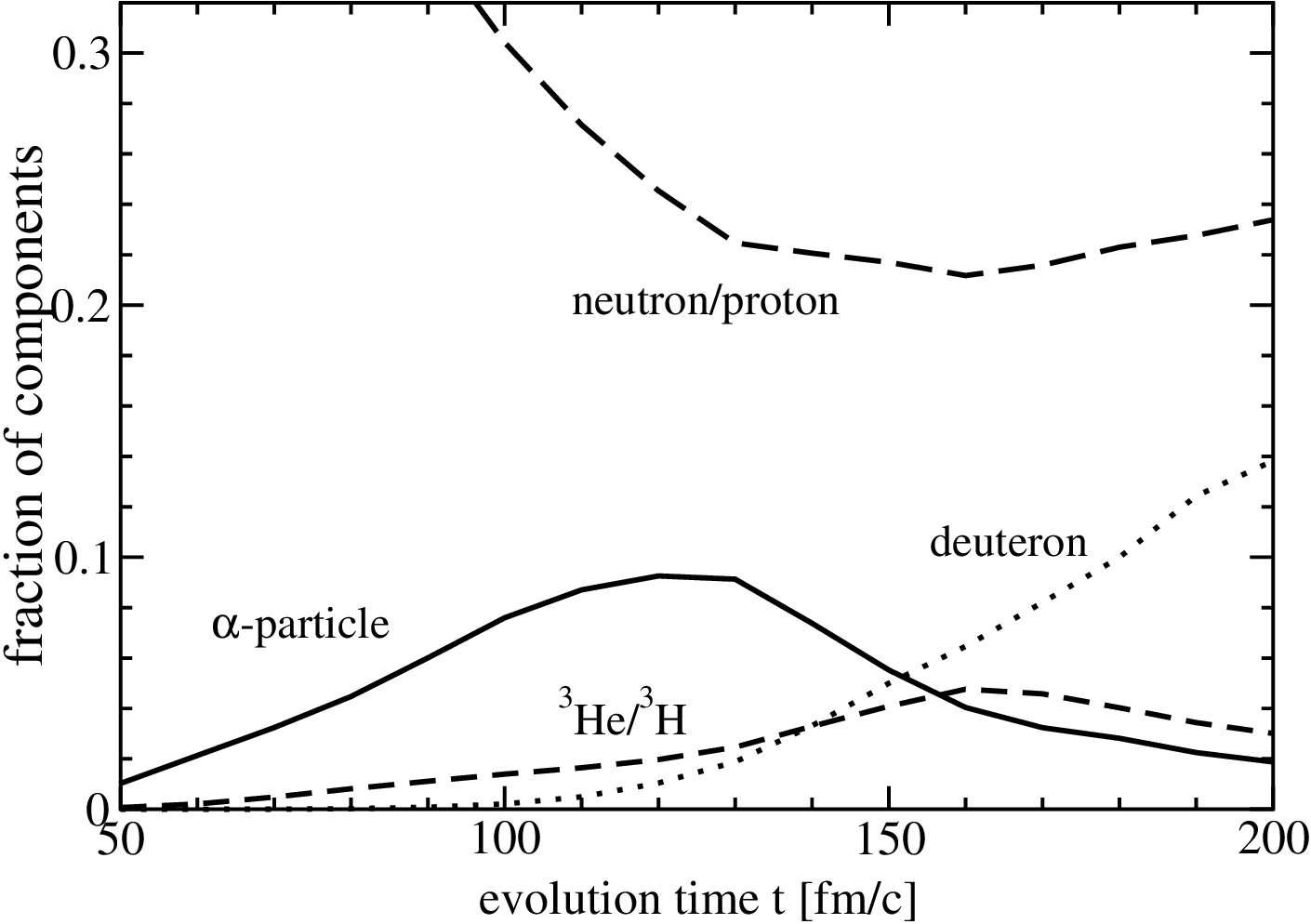}
~\\[10pt]\parbox{0.95\textwidth}
{\footnotesize {\bf FIGURE 1.} Fraction of protons, neutrons (both identical
  in symmetric nuclear matter, long-dashed), deuteron (dots), triton, $^3$He
  (both dashed), and $\alpha$ particles (solid) as a function of collision
  time at $T=10$ MeV.\vfill}
\label{fig1}
\end{minipage}\hfill
\begin{minipage}{0.5\textwidth}
\centering
\includegraphics[width=\textwidth]{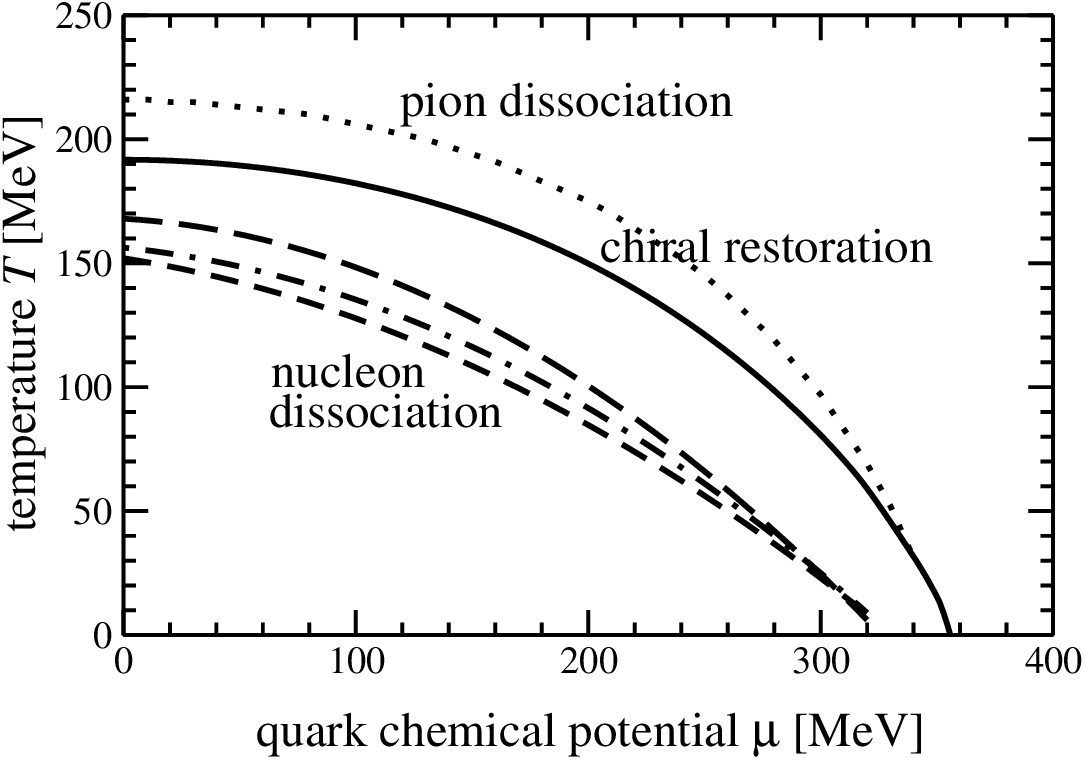}
\parbox{0.95\textwidth}
{\footnotesize~\\ {\bf FIGURE 1.} Phase diagram of QCD, chiral restoration (solid), 
pion dissociation line (dotted), nucleon dissociation 
lines for different cut-offs, $\Lambda=4m$ (long dashed), 
$6m$ (dashed-dotted), $8m$ (dahed), $m$ medium dependent quark mass.}
\label{fig2}
\end{minipage}
\end{figure}

\paragraph{Nuclear matter}
At lower densities and temperatures nuclear matter is composed out of nucleons
and nuclei.  To describe heavy ion collisions at energies of about 30-100 MeV
per nucleon it is sufficient to use a nonrelativistic treatment.  Typical
experiments have been performed by the INDRA collaboration~\cite{INDRA}. At
central collisions, where the system is likely to thermalize, the composition
shows a large multiplicities of $\alpha$-particles. This cannot be explained
by a simple ideal gas of components. The nuclei forming the gas are influenced
by the medium. The binding energies change and they can dissolve (Mott
transition). The equations follow from the Dyson expansion as explained in the
previous paragraph~\cite{Beyer:1996rx}.  In a simple fire ball expansion
(tailored to reproduce the Xe on Sn experiment at 50 MeVA~\cite{INDRA}) we assume a gas of
nuclei expanding during the heavy ion collision. This is shown in Fig.~1. At
around $t=100-120$ fm/c the number of $\alpha$-particles is much larger than
the number of other clusters.  This time, which corresponds to about 3-4 times
the initial volume, can be considered as freeze-out time. This freeze-out time
essentially determines the composition of nuclei hitting the detector. This
approach opens possibilities to address further questions related, e.g. to
the influence of three-nucleon correlations on superconductivity, to
four-nucleon condensation and even to larger clusters than $\alpha$-particles.

\paragraph{Quark matter - light front form}

At higher densities or temperatures we need a relativistic treatment. To this
end we have generalized the light-front quantization to statistical quantum
mechanics~\cite{Beyer:1996rx,Beyer:2001bc,Mattiello:2001vq}, see
also~\cite{LF}. The light-front quantization is capable to treat the
perturbative and the nonperturbative region on the same footing.  It can be
realized by using the Nambu Jona-Lasinio model as an example. This approach
reproduces the well known chiral phase transition, given as solid line in
Fig.~2. We have also calculated the pion (dotted) and the nucleon (several
dashed lines) dissociation line for different cut-off parameters.  One may
consider these lines related to a confinement-deconfinement transition, since
no bound states are possible above the respective lines.  The approach
provides the framework to tackle such intriguing questions how hadrons form
during the early plasma phase of the universe, if there is a color
superconducting phase, what is the nature of QCD phase transition. It is also
possible (and necessary) to go beyond the zero range model and implement real
light front QCD, which has been elaborated in the works by Brodsky, Pauli and
Pinsky~\cite{Brodsky:1997de} and is still progressing. The results are
promising and since we do have a light front form of QCD
available~\cite{Brodsky:1997de} it is exciting and challenging to extent this
theory to quantum statistics.

\paragraph{Acknowledgments}
  We would like to thank the organizers for a pleasant and inspiring
  atmosphere during this meeting. Work supported by Deutsche
  Forschungsgemeinschaft.



\bibliographystyle{aipproc}   

\end{document}